\documentclass[11pt]{article}
\usepackage[margin=1in]{geometry}
\usepackage{newtxtext,newtxmath} 

\usepackage{authblk}
\usepackage{graphicx}
\usepackage{titlesec}
\usepackage[colorlinks=false, pdfborder={0 0 0}]{hyperref}

\title{\vspace{-2em}Net 3.2 Tbps 225 Gbaud PAM4 O-Band IM/DD 2 km Transmission Using FR8 and DR8 with a CMOS 3 nm SerDes and TFLN Modulators}
\author[1]{Charles St-Arnault}
\author[1]{Santiago Bernal}
\author[3]{Derek Kita}
\author[2]{Ross Dickson}
\author[1]{Mariam Yehia Abdelaziz}
\author[1]{Aleksandar Nikic}
\author[1]{Benton Qiu}
\author[4]{Benjamin Krueger}
\author[4]{Fabio Pittalà}
\author[3]{Christian Reimer}
\author[2]{Bruce Beggs}
\author[2]{Naim Ben-Hamida}
\author[1]{David V. Plant}
\affil[1]{Dept. of Electrical and Computer Engineering, McGill University, Montreal, QC, Canada}
\affil[2]{Ciena Corporation, Ottawa, ON, Canada}
\affil[3]{HyperLight Corporation, Cambridge, MA, USA}
\affil[4]{Keysight Technologies Deutschland GmbH, Böblingen, Germany}
\date{}

\begin{document}

\maketitle

\vspace{-10mm}

\begin{center}
\texttt{charles.st-arnault@mail.mcgill.ca} \\
\vspace{1em}
\end{center}

\vspace{-4mm}
\begin{abstract}
We report the first 3.2 and 4.2 Tbps (8$\times$225Gbaud PAM4-8), IM/DD transmission system using FR8 and DR8 configurations with TFLN modulators driven by a 3nm SerDes under the HD-FEC threshold.
\end{abstract}

\vspace{-5mm}

\section{Introduction}
The IEEE 802.3 standard significantly evolved over the past 50+ years. In the last 5 years, more transmission speed standards were introduced than in the preceding 35 years \cite{ieee_ethernet_50th}. This acceleration is driven by the growing demand for higher data rates caused by AI data center mega projects. These data centers require scalable optics with low power consumption, high reliability, and increased data rates. To achieve this, power consumption can be improved by reducing DSP and transceiver complexity, utilizing more advanced CMOS nodes, and removing the TEC, enabled by uncooled DFB lasers. Such lasers can be used if the wavelength division multiplexing (WDM) grid can tolerate sizable laser wavelength drift, as is the case with the 20 nm spaced coarse-WDM (CWDM), or if a parallel single-mode fiber (SMF) transmission architecture is used (e.g., DR8). Due to chromatic dispersion (CD) and its impact on the expanding signaling bandwidths (e.g., 224 Gbaud), WDM grids are increasingly confined to narrower channel spacings, rendering the use of uncooled DFB lasers impractical, especially for distances larger than 2km \cite{JLT_16T}. In contrast, parallel SMF transmission architectures can use 1 or 2 lasers for 8 channels operating near the zero dispersion wavelength, minimizing the impacts of CD while improving transceiver reliability \cite{laserReliability}, and can operate uncooled, reducing power consumption. Last year, we demonstrated 1.6 Tbps at 2km using 160 Gbaud PAM8 on four wavelengths (1.6TBASE-FR4) \cite{PDP24}. This year, enabled by DAC and thin-film lithium niobate (TFLN) Mach Zehnder modulator (MZM) advancements, we demonstrate 3.2 and 4.2 Tbps with 225 Gbaud PAM4 and PAM8 using 8 wavelengths (3.2TBASE-FR8) at 2km and 8 parallel SMF channels (3.2TBASE-DR8+) at 500m and 2km. For the 8-WDM demonstration, we use an 8-channel 400 GHz spaced WDM grid starting and ending at 1295.56 and 1311.43 nm, respectively. For the DR8 demonstration, we use a 23 dBm, 1310 nm uncooled quantum dot (QD) DFB laser feeding 2$\times$4 channels, improving reliability and power consumption. We evaluate the performance of feed-forward equalizers (FFE), decision-feedback equalizers (DFE), and maximal-likelihood sequence estimation (MLSE) for the 448 Gbps/lane standards to address impairments such as CD and bandwidth (BW) limitations, as discussed by the 800G MSA group \cite{DiChe}.\\

\begin{figure}[t!]
\centering
\includegraphics[width=0.6\linewidth]{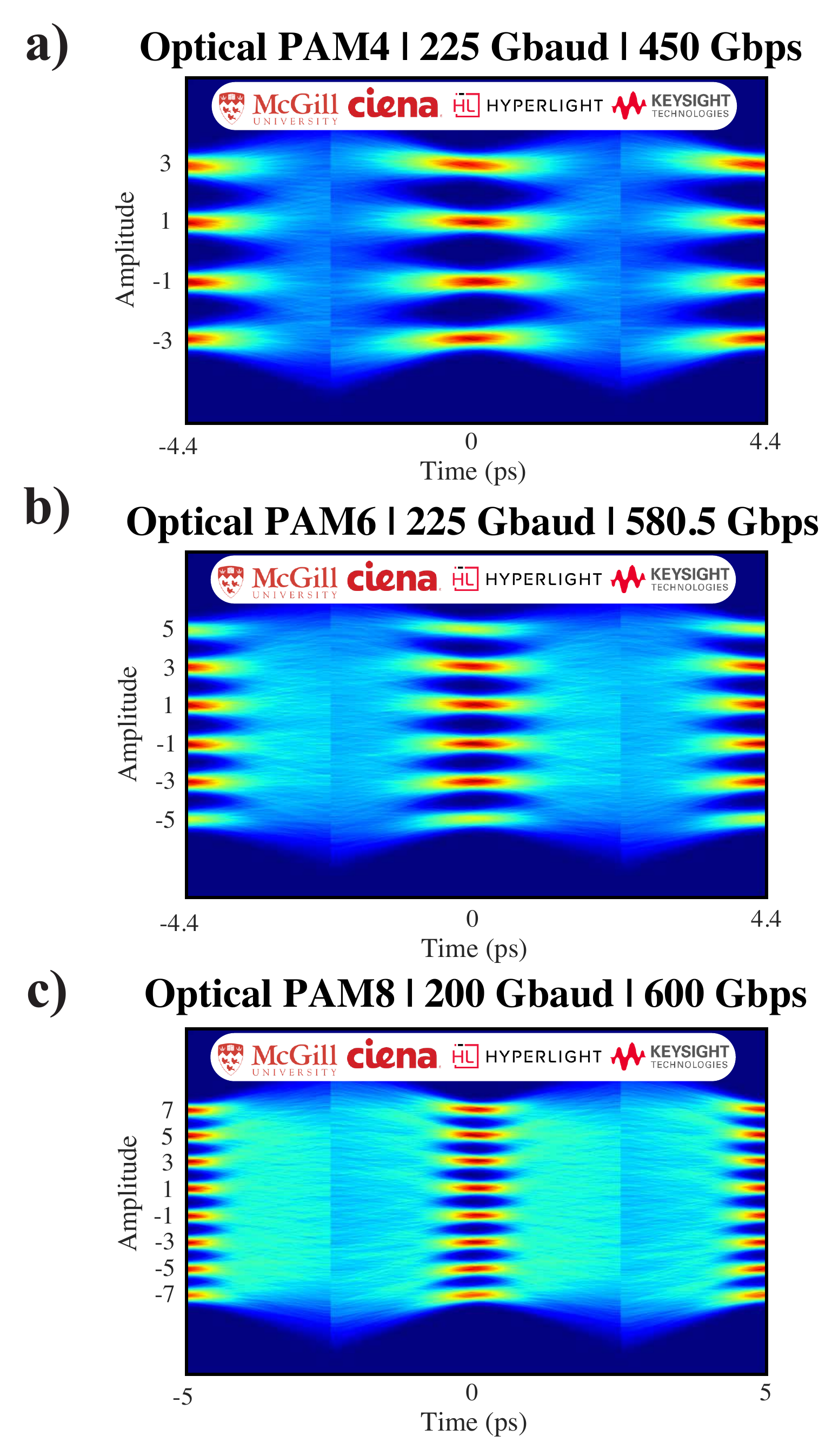}
\vspace{-4.5mm}
\caption{(a)-(c) PAM4-6-8 eye diagrams measured on a Keysight DCA with no waveform averaging, a 4th order Bessel-Thompson filter with half symbol rate cutoff frequency and 25 taps of FFE with 21 taps of DFE.}
\label{fig:eyes}
\end{figure}

Fig. \ref{fig:eyes} shows optical 225 Gbaud PAM4, 225 Gbaud PAM6 and 200 Gbaud PAM8 eye diagrams measured with a Keysight DCA. These eyes are captured with no waveform averaging, 25 taps of FFE and 21 taps of DFE along with a 4th order Bessel-Thompson filter with a cutoff frequency of half the symbol rate. The sharp transitions between eye openings are caused by the T-spaced DFE symbol decisions. These eye diagrams were captured with the experimental setup shown in Fig. \ref{fig:setup} (a).

\begin{figure}[t!]
\centering
\includegraphics[width=0.57\linewidth]{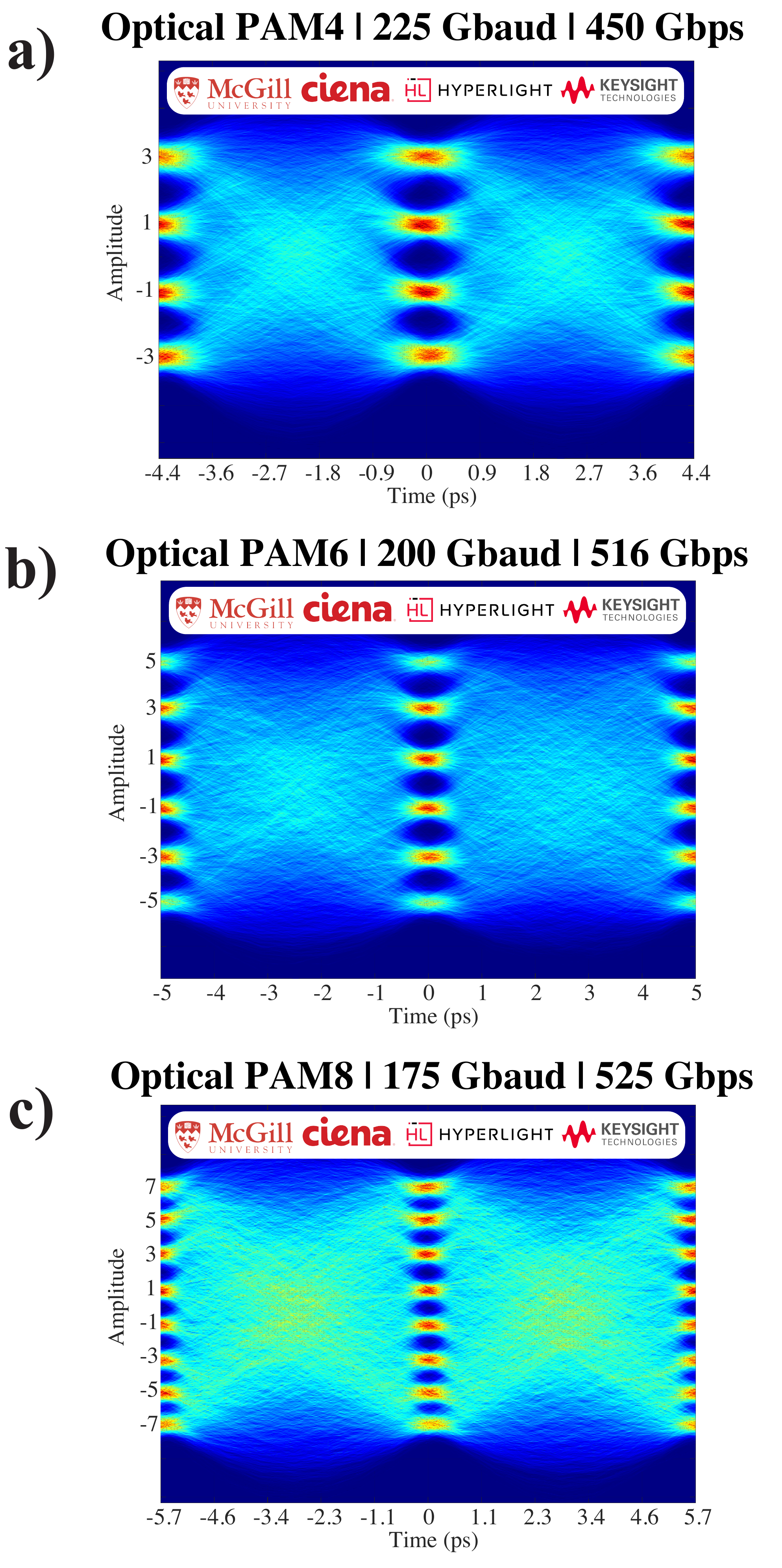}
\vspace{-4.5mm}
\caption{(a)-(c) PAM4-6-8 eye diagrams measured on a Keysight real-time oscilloscope (RTO) with 10 waveform averages and 51 taps of FFE with 21 taps of DFE.}
\label{fig:eyesRTO}
\end{figure}
\newpage
Fig. \ref{fig:eyesRTO} shows optical 225 Gbaud PAM4, 200 Gbaud PAM6 and 175 Gbaud PAM8 eye diagrams measured with a Keysight real-time oscilloscope (RTO). These eyes are captured with 10 waveform averages, 51 taps of FFE and 21 taps of DFE. These eye diagrams were captured with the experimental setup shown in Fig. \ref{fig:setup} (a).

\section{Experimental Setup}

\begin{figure}[t!]
\centering
\includegraphics[width=1\linewidth]{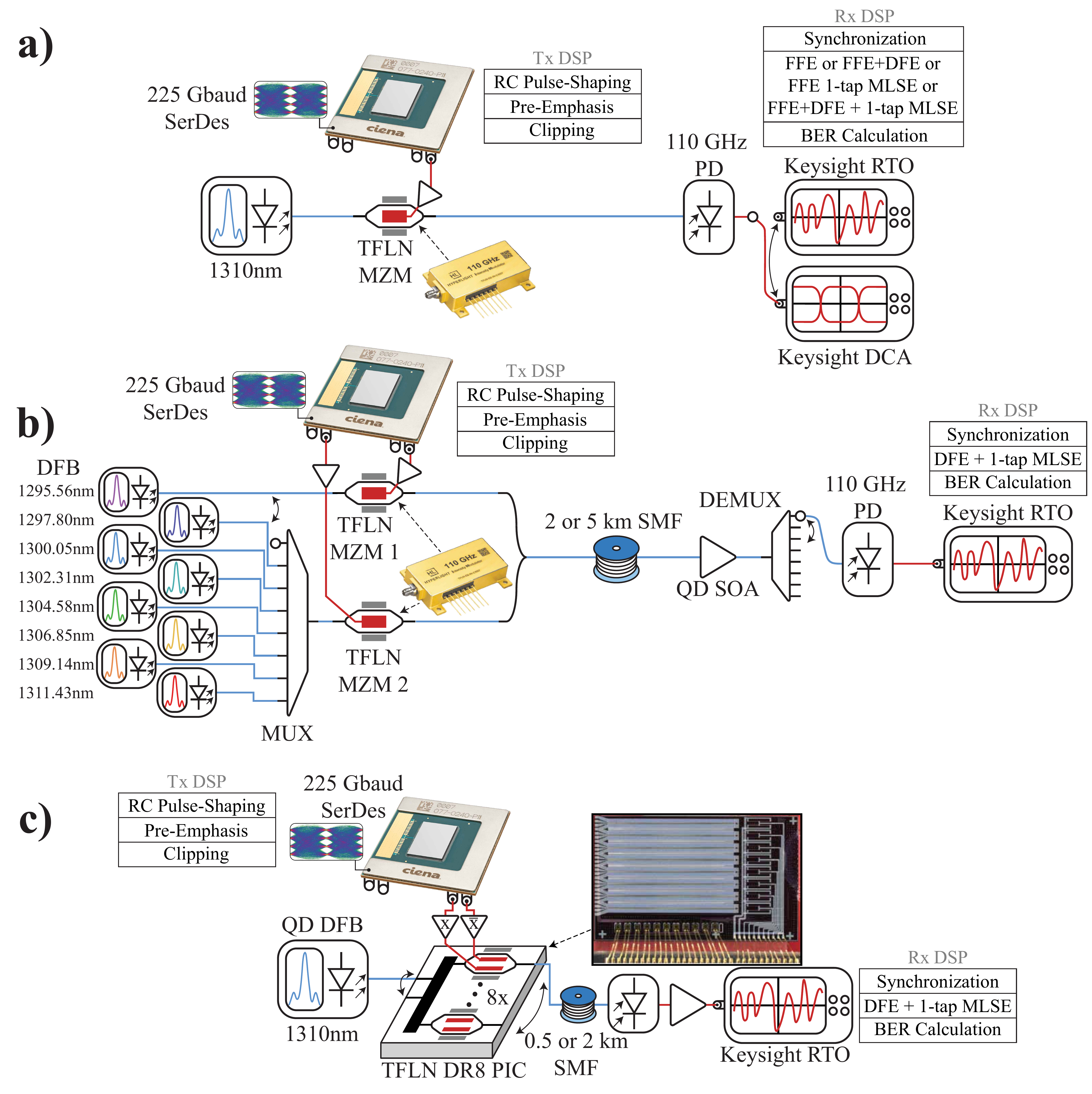}
\vspace{-8.5mm}
\caption{(a) The single wavelength experimental setup used to empirically compare receiver equalizer performance and capture eye diagrams. (b) The 8-WDM experimental setup used to emulate the 8 channel system. As shown, the channel under test (CUT ; 1295.56 nm) is modulated separately from the remaining 7 channels. Data collection is achieved by iterating the CUT through all 8 channels. (b) The DR8 experimental setup. A TFLN PIC containing 8 parallel MZMs is used. Data collection is achieved by iterating through all 8 fiber outputs. (c) Received optical 225 Gbaud PAM4 eye diagram with DFE as Rx equalizer.}
\label{fig:setup}
\end{figure}

Fig. \ref{fig:setup} (b) shows the experimental setup and DSP stacks used in the 8-WDM experiment. We generate $2^{18}$ symbols by Marsenne Twister, resample them to match the 2-channel 3 nm CMOS SerDes's 225 GSa/s sampling rate and apply pre-emphasis to the signal to compensate for the RF connections and the SerDes's DAC frequency response. This DAC has a 6 dB BW $>$ 110 GHz, a 7-bit resolution, a differential output voltage of 600 m$V_{pp}$ and inductive peaking at 100 GHz. Clipping is applied to optimize the peak to average power ratio (PAPR). Both PAM4(8)-coded signals are digitally delayed to decorrelate the symbols. Two packaged single-ended TFLN MZMs featuring a $>$ 110 GHz 3 dB electro-optic bandwidth and 2 V low-MHz $V_\pi$ are used to best emulate a real 8-WDM system. The first driving signal is amplified by a 105 GHz 3 dB BW Anritsu AH15199B RF amplifier and modulates the channel under test (CUT), while the second driving signal is amplified by a 100 GHz 3 dB BW SHF T850C RF amplifier and bulk modulates the combined remaining 7 wavelengths. As shown on the left of Fig. \ref{fig:setup} (a), eight 9 dBm DFB lasers spectrally arranged on a 400 GHz spaced WDM grid spanning from 1295.56 nm to 1311.43 nm are used. The modulated optical signals from MZM1 (CUT) and MZM2 are combined onto a single fiber using a 50/50 optical coupler. The CUT is iterated through all 8 channels, and at all times, 8 modulated signals are propagating through the fiber. The combined optical signals then travel through 2 or 5km of SMF. A QD SOA \cite{QDSOA} is used to compensate for excessive optical losses, reduce the laser launch power requirements and eliminates the need for an RF driver or TIA after the photo-detector (PD). The CUT is selected using a demultiplexer and converted into an electrical signal by a 110 GHz PD. The received signal is then sampled by a 110 GHz Keysight real-time oscilloscope (RTO), resampled to 2 samples per symbol and synchronized. Next, a DFE followed by a 1-tap MLSE is applied. The equalized symbols are then mapped back to a binary sequence and compared to the transmitted sequence to calculate the BER. Fig. \ref{fig:setup} (a) shows a simplified version of the 8-WDM experiment where a single wavelength is used. This setup is used to empirically evaluate the performance of FFE, DFE and both with 1-tap of MLSE added. No optical amplifier is used in this setup. Fig. \ref{fig:setup} (c) shows the DR8 experimental setup. The same Tx and Rx DSP stacks are used. A 23 dBm QD DFB laser feeds all 8 ($>$ 110 GHz 3 dB BW) 4.5 V differential $V_\pi$ MZMs on the DR8 PIC via two 1x4 power splitter MMI circuits. The TFLN DR8 PIC features 11 bonded monitor PDs. Each MZM is modulated one at a time by a differential output from the DAC and amplified by two SHF T850C amplifiers. Crosstalk between adjacent MZMs was measured, and was not detected on the RTO. The measured channel is selected by choosing the appropriate output fiber and measurements are repeated on all 8 modulators. The modulated optical signal then travels through 2km of SMF and is converted to an electrical signal by a 110 GHz PD. This electrical signal is amplified by an Anritsu AH15199B amplifier and sampled by the Keysight RTO.\\

\newpage
\newpage
\section{Results}

\begin{figure}[t]
\centering
\includegraphics[width=1.009\linewidth]{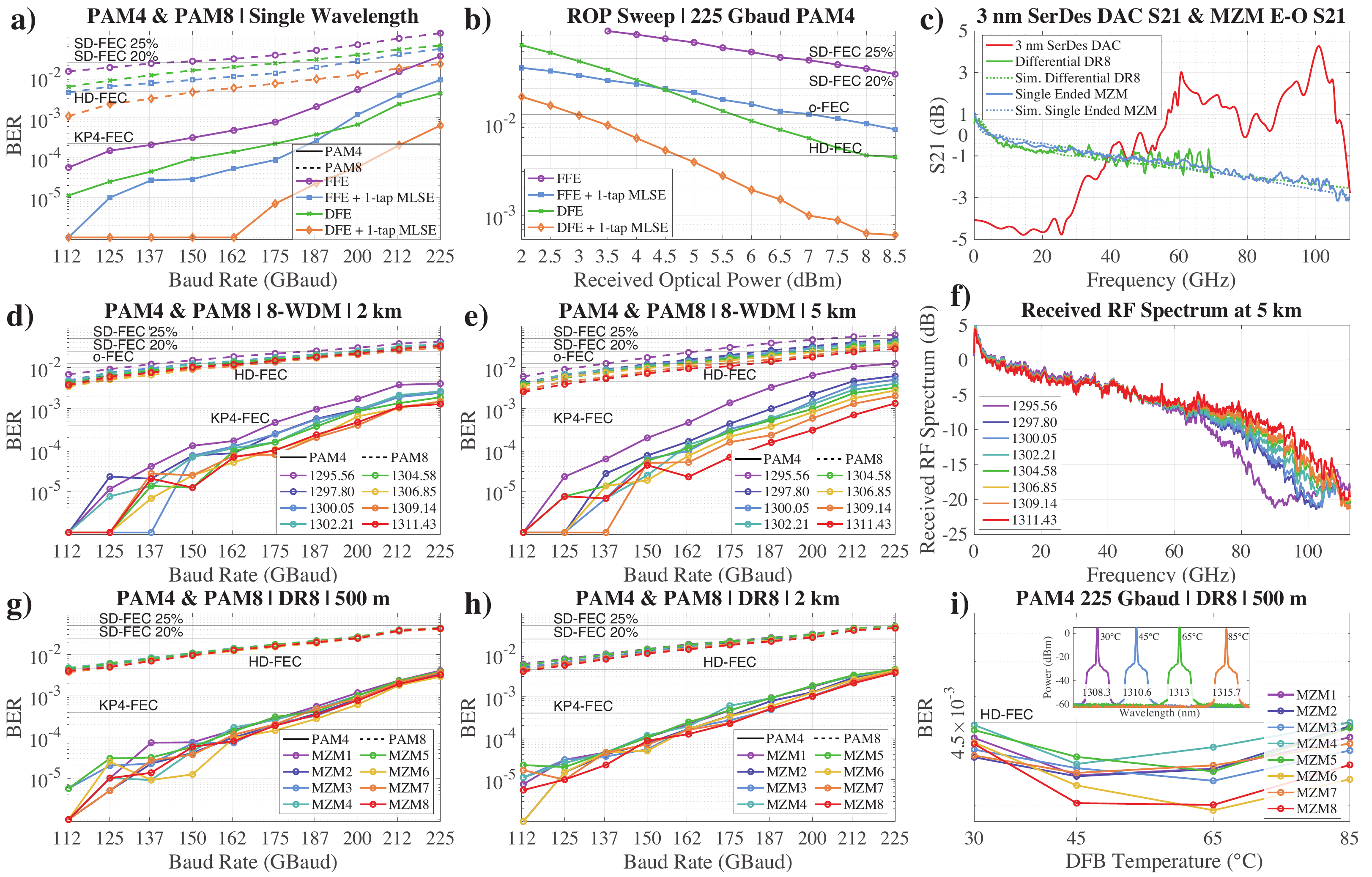}
\vspace{-7.5mm}
\caption{(a) BER vs symbol rate for single wavelength PAM4 and PAM8 at back-to-back distance comparing FFE, FFE + 1-tap MLSE, DFE and DFE + 1-tap MLSE receiver equalizers. (b) BER vs received optical power (ROP) for 225 Gbaud PAM4 with the 4 tested receiver equalizers. (c) S21 of the 3 nm SerDes DAC, differential DR8 and single ended MZMs. (d)-(e) BER vs symbol rate for 8-WDM PAM4 and PAM8 at (d) 2km and (e) 5km. (f) Received RF spectrum for the 8-WDM transmission at 5km showing CD-induced power fading and dips. (g)-(h) BER vs symbol rate for DR8 PAM4 and PAM8 at (g) 500m and (h) 2km. (i) BER vs uncooled DFB laser temperature for 225 Gbaud PAM4 at 500m.}
\label{fig:results}
\end{figure}

Fig. \ref{fig:results} (a)-(c) contain the single wavelength experimental results obtained with the setup in Fig. \ref{fig:setup} (a) without multiplexers and optical amplifier. Fig. \ref{fig:results} (a) presents the PAM4 (solid) and PAM8 (dashed) results for 4 receiver equalizers: FFE, DFE and each combined with 1-tap of MLSE. Non-linear symbol decision and sequential based equalizers were selected to combat inter-symbol interference caused by BW limitations and the colored noise in our system, where the noise PSD of the Keysight RTO peaks at 110 GHz. Linear equalizers used on signals with high BW limitations, shown in Fig. \ref{fig:results} (f), further exacerbate noise coloration, degrading transmission performance, as shown in Fig. \ref{fig:results} (a) where the FFE has a BER 8.5x higher than the DFE for 225 Gbaud PAM4. Adding a single tap of MLSE to the FFE reduces the BER by 4x. For symbol rates $>$ 187 Gbaud, the DFE performs better than FFE + 1-tap MLSE. At 225 Gbaud PAM4, the DFE equalizer achieves a BER under the 7\% overhead (OH) HD-FEC threshold giving a net rate of 420.5 Gbps. Adding one tap of MLSE to the DFE reduced the BER by 6.4x at symbol rates greater than 200 Gbaud, enabling 212.5 Gbaud PAM4 under the 5.8\% OH KP4-FEC, for a net rate of 401.7 Gbps. Similar trends in equalizer performance are observed for PAM8, although, stricter SNR requirements make DFE perform worse when compared to FFE + 1-tap MLSE due to error propagation sensitivity. Notwithstanding, when paired with 1-tap MLSE, the DFE achieves 225 Gbaud PAM8 under the 20\% OH SD-FEC giving a net rate of 562.5 Gbps. Fig. \ref{fig:results} (b) shows the BER versus received optical power (ROP) for 225 Gbaud PAM4. The BER slopes for the DFE-based equalization schemes show a larger slope with decreasing ROP indicating the error propagation sensitivity of DFE equalizers. At $\leq$4.5 dBm, the FFE + 1-tap MLSE begins to outperform the DFE. Considering these results, the DFE + 1-tap MLSE was the chosen receiver equalizer for the 8-channel experiments. Fig. \ref{fig:setup} (c) shows an optical 225 Gbaud PAM4 eye diagram using DFE and 10 waveform averages. Fig. \ref{fig:results} (d) shows the PAM4 and PAM8 8-WDM results at 2km. For both PAM formats, the edge-most channel (1295.56 nm) suffers the most CD and thus performs worse. All 8 wavelengths achieve net 420.5 Gbps with 225 Gbaud PAM4 under the 7\% HD-FEC threshold giving an aggregate rate of 3.36 Tbps. For PAM8, all wavelengths achieve net rates of 540 Gbps with 225 Gbaud PAM8 under the 25\% OH SD-FEC threshold giving an aggregate rate of 4.32 Tbps. At 5km, the accumulated CD on the edge channels, as shown in Fig. \ref{fig:results} (f), increase the BER for 225 Gbaud PAM4 above the HD-FEC. Assuming the same FEC threshold for all channels, the best achieved net rate is 375 Gbps with 225 Gbaud PAM4 under the 20\% SD-FEC threshold giving an combined rate of 3.0 Tbps. Using PAM8, a combined rate of 3.84 Tbps is achieved with 200 Gbaud PAM8 under the 25\% SD-FEC threshold.

Fig. \ref{fig:results} (g) and (h) show the PAM4 and PAM8 DR8 results at 500m and 2km respectively. With 225 Gbaud PAM4, net 420.5 Gbps was achieved under the 7\% OH HD-FEC threshold, giving an aggregate rate of 3.36 Tbps for both distances. These results were taken with the laser temperature at 40°C. With 225 Gbaud PAM8, net 540 Gbps was achieved under the 25\% OH SD-FEC threshold yielding an aggregate rate of 4.32 Tbps for both distances. Fig. \ref{fig:results} (i) shows the BER versus DFB laser temperature at 225 Gbaud PAM4 at 500m for 30°C to 85°C. At all temperatures, the BER remains under the HD-FEC, showing uncooled operation across a 55°C range. At 30°C, the wavelength is 1308.3 nm and at 85°C, the wavelength is red shifted to 1315.7 nm complying with the 200 Gbps/lane DR8 carrier wavelength specification \cite{IEEE_DR8}. BER fluctuations are caused by the laser power varying by 0.8 dBm.

\begin{table}[]
    \centering
    \caption{Summary of achieved net data rates}
\begin{tabular}{|ccccc|}
\hline
\multicolumn{1}{|c|}{\begin{tabular}[c]{@{}c@{}}\textbf{BER} \\      \textbf{Threshold}\end{tabular}} & \multicolumn{1}{c|}{\begin{tabular}[c]{@{}c@{}}\textbf{FEC}\\      \textbf{Overhead}\end{tabular}} & \multicolumn{1}{c|}{\begin{tabular}[c]{@{}c@{}}\textbf{Symbol Rate /} \\       \textbf{Modulation Format}\end{tabular}} & \multicolumn{1}{c|}{\begin{tabular}[c]{@{}c@{}}\textbf{Net Rate}\\      \textbf{(Gbps)}\end{tabular}} & \multicolumn{1}{c|}{\begin{tabular}[c]{@{}c@{}} \textbf{Aggregate Rate}\\      \textbf{(Tbps)}\end{tabular}} \\ \hline \hline

\multicolumn{5}{|c|}{\textbf{Single Wavelength (1310 nm) B2B}}                                                                                                                                                                                                                                                                                                                                                                                                                                                                                                \\ \hline
\multicolumn{1}{|c|}{$4.5 \times 10^{-3}$}                                    & \multicolumn{1}{c|}{7\%}                                                        & \multicolumn{1}{c|}{225 Gbaud / PAM4}                                                            & \multicolumn{1}{c|}{420.5}                                                        & \multicolumn{1}{c|}{-}                                                                  \\ \hline
\multicolumn{1}{|c|}{$2.4 \times 10^{-2}$}                                      & \multicolumn{1}{c|}{20\%}                                                        & \multicolumn{1}{c|}{225 Gbaud / PAM8}                                                            & \multicolumn{1}{c|}{562.5}                                                        & \multicolumn{1}{c|}{-}                                                                   \\ \hline \hline

\multicolumn{5}{|c|}{\textbf{DR8 (500m), DR8+ and 8-WDM (2km)}}                                                                                                                                                                                                                                                                 \\ \hline
\multicolumn{1}{|c|}{$4.5 \times 10^{-3}$}                                    & \multicolumn{1}{c|}{7\%}                                                        & \multicolumn{1}{c|}{225 Gbaud / PAM4}                                                            & \multicolumn{1}{c|}{8 $\times$ 420.5}                                                        & \multicolumn{1}{c|}{3.36}                                                                  \\ \hline
\multicolumn{1}{|c|}{$5 \times 10^{-2}$}                                      & \multicolumn{1}{c|}{25\%}                                                        & \multicolumn{1}{c|}{225 Gbaud / PAM8}                                                            & \multicolumn{1}{c|}{8 $\times$ 540}                                                        & \multicolumn{1}{c|}{4.32}                                                                   \\ \hline \hline

\multicolumn{5}{|c|}{\textbf{8-WDM (5km)}}                                                                                                                                                                                                                                                                 \\ \hline
\multicolumn{1}{|c|}{$2.4 \times 10^{-2}$}                                    & \multicolumn{1}{c|}{20\%}                                                        & \multicolumn{1}{c|}{225 Gbaud / PAM4}                                                            & \multicolumn{1}{c|}{8 $\times$ 375}                                                        & \multicolumn{1}{c|}{3.0}                                                                  \\ \hline
\multicolumn{1}{|c|}{$5 \times 10^{-2}$}                                      & \multicolumn{1}{c|}{25\%}                                                        & \multicolumn{1}{c|}{200 Gbaud / PAM8}                                                            & \multicolumn{1}{c|}{8 $\times$ 480}                                                        & \multicolumn{1}{c|}{3.84}                                                                   \\ \hline
\end{tabular}
\label{tab:results}
\end{table}

\section{Conclusion}

A net 3.2-4.2 Tbps IM/DD transmission over 2km - using 225 Gbaud PAM4 (450 Gbps) and 225 Gbaud PAM8 (675 Gbps) - was demonstrated for both 8-WDM and DR8 configurations.

\bibliographystyle{ieeetr}
\bibliography{references}

\begin{thebibliography}{1}

\bibitem{ieee_ethernet_50th}
{IEEE Standards Association}, ``{Ethernet Through the Years: Celebrating the Technology’s 50th Anniversary}.''

\bibitem{JLT_16T}
C.~St-Arnault, R.~Gutiérrez-Castrejón, S.~Bernal, E.~Berikaa, Z.~Wei, J.~Zhang, M.~S. Alam, A.~Nikic, B.~Qiu, B.~Krueger, F.~Pittalà, and D.~V. Plant, ``{Practical Fiber Dispersion-Induced Limitations for 1.6 Tbps (4× 400 Gbps/$\lambda$) O-Band IM/DD Transmission Systems Over 2, 10, 20 and 40 km},'' {\em Journal of Lightwave Technology}, vol.~43, no.~7, pp.~3222--3232, 2025.

\bibitem{laserReliability}
C.~Xie, C.~Wang, Q.~Chen, Z.~Wang, P.~Wang, R.~Lu, and L.~Wang, ``{Characteristics of Field Operation Data for Optical Transceivers in Hyperscale Data Centers},'' in {\em 2022 Optical Fiber Communications Conference and Exhibition (OFC)}, pp.~01--03, 2022.

\bibitem{PDP24}
C.~St-Arnault, S.~Bernal, E.~Berikaa, Z.~Wei, R.~Gutiérrez-Castrejón, J.~Zhang, M.~S. Alam, A.~Nikic, B.~Krueger, F.~Pittalà, and D.~V. Plant, ``{Net 1.6 Tbps (4×400Gbps/$\lambda$) O-Band IM/DD Transmission Over 2 km Using Uncooled DFB Lasers on the LAN-WDM grid and Sub-1V Drive TFLN Modulators},'' in {\em 2024 Optical Fiber Communications Conference and Exhibition (OFC)}, pp.~1--3, 2024.

\bibitem{DiChe}
D.~Che and X.~Chen, ``{Modulation Format and Digital Signal Processing for IM-DD Optics at Post-200G Era},'' {\em Journal of Lightwave Technology}, vol.~42, no.~2, pp.~588--605, 2024.

\bibitem{QDSOA}
C.~St-Arnault, S.~Bernal, R.~Gutiérrez-Castrejn, E.~Berikaa, W.~Li, Z.~Wei, Y.~Hu, M.~S. Alam, J.~Rautert, S.~V. Poltavtsev, A.~E. Gubenko, V.~V. Belykh, V.~S. Mikhrin, A.~R. Kovsh, and D.~V. Plant, ``{Performance and Characterization Comparison of QD SOA, QW SOA, Bulk SOA and PDFA for Multi-Tbps O-Band WDM Links},'' {\em Journal of Lightwave Technology}, vol.~43, no.~4, pp.~1915--1925, 2025.

\bibitem{IEEE_DR8}
{IEEE P802.3dj 200 Gb/s, 400 Gb/s, 800 Gb/s, and 1.6 Tb/s Ethernet Task Force}, ``{Baseline proposals for 200G/L PMD specifications for single wavelength 500m and 2km standards}.''

\end{thebibliography}

\end{document}